\documentclass{aipproc}

\usepackage{epsfig} 
\usepackage{amsbsy} 

\def\beq{\begin{equation}}
\def\eeq{\end{equation}}

\def\beqn{\begin{eqnarray}}
\def\eeqn{\end{eqnarray}}

\def\lq{\left[}
\def\rq{\right]}

\def\({\left(}
\def\){\right)}

\newcommand{\n}[1]{\nu_{#1}}

\def\xbox{{\rm Xbox}}

\def\xbm#1#2#3#4{\xbox_{#1}^{#2}\(#3,#4\)}

\def\tm1#1#2{{\rm Btri}^{#1}\(#2\)}

\def\IBP{{\rm integration-by-parts}}

\def\f21{{_2F_1}}

\def\so3#1{\,{\rm S}_{1,\,3}\left(#1 \right)}
\def\st2#1{\,{\rm S}_{2,\,2}\left(#1 \right)}

\newcommand{\Li}[2]{{\mbox{Li}}_{#1}\left(#2\right)}
\newcommand{\snp}[2]{{\mbox{S}}_{#1}\left(#2\right)}
\def\mtos{-\frac{t}{s}}

\def \ep{\epsilon}
\def \to   {\mbox{$\rightarrow$}}

\newcommand\hepph[1]{{\tt hep-ph/#1}}

\newcount\minutes
\newcount\scratch

\def\timestamp{%
\scratch=\time
\divide\scratch by 60
\edef\hours{\the\scratch}
\multiply\scratch by 60
\minutes=\time
\advance\minutes by -\scratch
---$\,$\hours:\null
\ifnum\minutes< 10 0\fi
\the\minutes}

\newcommand{\SUNC}[1]{
\mbox{\parbox{3cm}{
\begin{picture}(3.5,1.4)
\thicklines
\put(0.5,0.7){\line(1,0){2}}
\put(1.5,0.7){\circle{1}}
\put(2.65,0.7){\makebox(0,0)[l]{$(#1)$}}
\end{picture}
}}
\hfill}

\newcommand{\TRI}[1]{
\mbox{\parbox{3cm}{
\begin{picture}(3.5,1.4)
\thicklines
\put(0.5,0.7){\line(1,0){0.5}}
\put(1.5,1.2){\line(0,-1){1}}
\put(1.5,1.2){\line(1,0){1}}
\put(1.5,0.2){\line(1,0){1}}
\put(1.5,0.7){\circle{1}}
\put(2.65,0.7){\makebox(0,0)[l]{$(#1)$}}
\end{picture}
}}
\hfill}

\newcommand{\ABOX}[2]{
\mbox{\parbox{3cm}{
\begin{picture}(3.5,1.4)
\thicklines
\put(0.5,0.2){\line(1,0){2.4}}
\put(0.5,1.2){\line(1,0){2.4}}
\put(1,0.2){\line(0,1){1}}
\put(2,0.7){\circle{1}}
\put(3,0.7){\makebox(0,0)[l]{$(#1,#2)$}}
\end{picture}
}}
\hfill}

\newcommand{\CBOX}[2]{
\mbox{\parbox{3cm}{
\begin{picture}(3.5,1.4)
\thicklines
\put(0.5,0.2){\line(1,0){2.4}}
\put(1.2,0.2){\line(1,1){1}}
\put(0.5,1.2){\line(1,0){2.4}}
\put(1.2,0.2){\line(0,1){1}}
\put(2.2,0.2){\line(0,1){1}}
\put(3.05,0.7){\makebox(0,0)[l]{$(#1,#2)$}}
\end{picture}
}} 
\hfill}

\newcommand{\Xboxa}[2]{
\mbox{\parbox{3cm}{
\begin{picture}(3.5,1.4)
\thicklines
\put(0.5,0.2){\line(1,0){2.4}}
\put(0.5,1.2){\line(1,0){2.4}}
\put(1,0.2){\line(0,1){1}}
\put(1.7,0.2){\line(1,1){1}}
\put(1.7,1.2){\line(1,-1){1}}
\put(3.05,0.7){\makebox(0,0)[l]{$(#1,#2)$}}
\end{picture}
}} 
\hfill}

\newcommand{\Xboxb}[2]{
\mbox{\parbox{3cm}{
\begin{picture}(3.5,1.4)
\thicklines
\put(0.5,0.2){\line(1,0){2.4}}
\put(1.7,0.2){\line(1,1){1}}
\put(0.5,1.2){\line(1,0){2.4}}
\put(1,0.2){\line(0,1){1}}
\put(1.7,1.2){\line(1,-1){1}}
\put(1,0.7){\circle*{0.2}}
\put(3.05,0.7){\makebox(0,0)[l]{$(#1,#2)$}}
\end{picture}
}} 
\hfill}

\newcommand{\Xtri}[1]{
\mbox{\parbox{3cm}{
\begin{picture}(3.5,1.4)
\thicklines
\put(0.5,0.7){\line(1,0){0.7}}
\put(1.2,0.7){\line(3,1){2.2}}
\put(1.2,0.7){\line(3,-1){2.2}}
\put(1.8,0.9){\line(3,-2){1.2}}
\put(1.8,0.5){\line(3,2){1.2}}
\put(3.05,0.7){\makebox(0,0)[l]{$(#1)$}}
\end{picture}
}} 
\hfill}

\begin{document}

\title{The tensor reduction and master integrals  of the two-loop massless
crossed box\footnote{Talk given
at {\sl ACAT 2000}, Fermilab, Batavia, IL, October 16--20, 2000. To be
published in {\it Advanced Computing and Analysis Techniques in Physics
Research}, edited by: P.C.~Bhat and M.~Kasemann.}} 
\author{Carlo Oleari}
\affiliation{Department of Physics, University of Wisconsin, 1150 University
Avenue, Madison WI 53706, U.S.A.}

\begin{abstract}
We briefly discuss an algorithm for the tensor
reduction of the two-loop massless crossed boxes, with light-like external
legs, and the computation of the relative master integrals.
\end{abstract}

\maketitle

\section{Introduction}
The level reached nowadays by the precision measurements in high-energy
scattering experiments demands  the knowledge of 
next-to-next-to-leading theoretical amplitudes  for $2\,\to\, 2$
scattering processes.

Very recent results for two-loop scattering amplitudes for massless particle
have already appeared 
in the literature: the maximal-helicity-violating two-loop
amplitude for $gg \to gg$~\cite{bdk}, $e^+e^- \to \mu^+\mu^-$ and $e^+e^- \to
e^-e^+$~\cite{BDG}, $q \bar q \to q^\prime \bar{q}^\prime $~\cite{qqQQ}
and $q \bar q \to q \bar{q} $~\cite{qqqq}.

In dealing with these two-loop scattering amplitudes we have to face the
problem of the tensor reduction of planar~\cite{Smirnov_Veretin} and crossed
double boxes~\cite{Xbox}, plus a plethora of simpler
topologies~\cite{pentabox,AGO2}, and the computation of the relative master
integrals~\cite{Smirnov,Bas}.

\section{Notation}
\label{sec:notation}

\begin{figure}[htb]
\centerline{\epsfig{figure=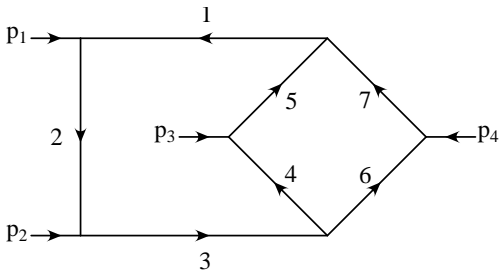,width=0.4\textwidth,clip=}}
\caption{The generic two-loop crossed  box.}
\label{fig:Xboxfig}
\end{figure}
We denote the generic two-loop tensor crossed (or non-planar) four-point
function in $D$ dimensions of Fig.~\ref{fig:Xboxfig} with seven propagators
$A_i$ raised to arbitrary powers $\n{i}$ as
\beqn
&&\xbox^D\(\nu_1\ldots\nu_7;s,t\)\lq 1;\, k^\mu;\, l^\mu;\, k^\mu k^\nu;\,k^\mu
l^\nu;\,\ldots \rq 
\nonumber\\
&&{}
=\int \frac{d^D k}{i \pi^{D/2}}
\int \frac{d^D l}{i \pi^{D/2}}
~
\frac{\lq 1;\, k^\mu;\, l^\mu;\, k^\mu k^\nu;\,k^\mu l^\nu;\,\ldots \rq}{
A_1^{\nu_1}
A_2^{\nu_2}
A_3^{\nu_3}
A_4^{\nu_4}
A_5^{\nu_5}
A_6^{\nu_6}
A_7^{\nu_7}},\nonumber
\eeqn
where the propagators are 
\[
\begin{array}{ll}
\label{eq:propagators}
A_1 = (k+l+p_{34})^2 + i0, & A_2 = (k+l+p_{134})^2 + i0, \nonumber \\
A_3 = (k+l)^2 + i0, & A_4 = l^2 + i0, \\
A_5 = (l+p_3)^2 + i0, & A_6 = k^2 + i0, \nonumber \\
A_7 = (k+p_4)^2 + i0. \nonumber
\end{array}\]
The external momenta $p_j$ are in-going and light-like, $p_j^2=0$,
$j=1\ldots 4$, so that the only momentum scales are the usual
Mandelstam variables $s = (p_1+p_2)^2$ and $t=(p_2+p_3)^2$, together with
$u=-s-t$.
For ease of notation, we define $p_{ij}=p_i+p_j$ and
$p_{ijk}=p_i+p_j+p_k$. 
In the square brackets we keep trace of the tensor
structure that may be  present in the numerator. 

\section{Tensor reduction}
\label{sec:tensor_redux}
As it is well known, tensor integrals can be related to combinations of
scalar integrals with higher powers of propagators and/or different values of
$D$~\cite{pentabox,Tar}.  
This is quite straightforward to see if we rewrite the Feynman integral
introducing the Schwinger parameters, diagonalizing the exponent of the
out-coming integral and integrating out the loop momenta~\cite{Xbox,pentabox}.

The task to compute tensor integrals is then moved to the
computation of scalar integrals with 
\begin{itemize}
\item[-] higher powers of the propagators 
\item[-] in higher dimensions.
\end{itemize}
From the Schwinger representation of Feynman integrals it easy to see that
integrals in $D$ dimensions can be connected with integrals in $D+2$
dimensions (dimensional-shift)~\cite{Smirnov_Veretin,Xbox}.

In this way, tensor integrals can be directly connected to scalar integrals
with higher powers of the propagators in $D=4-2\ep$ dimensions.

\section{The scalar crossed-box reduction}
\label{sec:reduction}
The strategy to reduce the generic scalar integral to a linear combination of
known ones is based on recurrence identities that relate scalar integrals with
different powers of propagators.  Some of these identities can be obtained
using the \IBP\ method~\cite{ibyp} and exploiting the Lorentz invariance of the
Feynman diagram~\cite{GR}.

Following the reduction procedure detailed in Ref.~\cite{Xbox},
any scalar crossed box with arbitrary powers of the propagators
can be written as a linear combination of the following integrals, that, 
therefore, are called master integrals:
\begin{eqnarray*}
{\rm Sset}^D(s) &=& \SUNC{s}\\
{\rm Tri}^D(s) &=&  \TRI{s} \\
{\rm Xtri}^D(s) &=& \Xtri{s} \\
{\rm Bbox}^D(s,t) &=& \ABOX{s}{t} \\
{\rm Dbox}^D(s,t) &=& \CBOX{s}{t} \\
\xbox_1^D(s,t) &=& \Xboxa{s}{t} \\
\xbox_2^D(s,t) &=&\Xboxb{s}{t},
\end{eqnarray*} 
where all the propagators have powers one except for the propagator with the
blob, that has power two.

\section{Differential equations for the two master integrals}
\label{sec:system}
The analytic expansion in $\ep=(4-D)/2$ for the first master cross box
$\xbox_1^D(s,t)$ was computed in Ref.~\cite{Bas}.  We can obtain the analytic
form for the second one by writing the derivative of $\xbox_1^D(s,t)$ with
respect to one of the two independent physical scales (that we choose to be
$t$), as a combination of master integrals, and solving the equation for
$\xbox_2^D(s,t)$.
Moreover we can verify the correctness of both the expressions of
$\xbm1{D}{s}{t}$ and $\xbm2{D}{s}{t}$, by deriving an analogous differential
equation for $\xbox_2^D$, and checking that the obtained identity is satisfied.
 
Starting from the Schwinger representation of the generic crossed box, we can
differentiate with respect to $t$, and set the values of $\nu_i$ to reproduce
the two master integrals 
\beqn
\frac{\partial}{\partial t} \xbm1{D}{s}{t} &=& 
\xbox^{D+2}\(1,2,1,2,1,1,2;s,t\)\nonumber\\
&&-\xbox^{D+2}\(1,2,1,1,2,2,1;s,t\),\nonumber\\
\frac{\partial}{\partial t} \xbm2{D}{s}{t} &=& 
2\, \xbox^{D+2}\(1,3,1,2,1,1,2;s,t\)\nonumber\\
&&- 2\,\xbox^{D+2}\(1,3,1,1,2,2,1;s,t\).
\nonumber
\eeqn
Applying the reduction formalism for the scalar integrals and the
dimensional-shift, we can rewrite the right-hand sides of the system as a
combination of the two master crossed boxes plus other master integrals of
simpler topologies, obtained by pinching one or more of the propagators of
the crossed box,
\beqn
\label{eq:der_t_xbm1}
&& \hspace{-0.5cm}
\frac{\partial}{\partial t} \xbm1{D}{s}{t} =
\frac{1}{t-u} \bigg[ \frac{(D-4) s^2-4 t u}{2 t u} \ \xbm1{D}{s}{t}\nonumber\\
&&\quad  
-\frac{(D-6) s}{2 (D-5)}\  \xbm2{D}{s}{t} + {\rm pinchings}\bigg] 
\\
\label{eq:der_t_xbm2}
&&\hspace{-0.5cm}
\frac{\partial }{\partial t} \xbm2{D}{s}{t}  =
\frac{1}{t-u}  \bigg[ \frac{2 (D-5)^2 s}{ t u}\  \xbm1{D}{s}{t} \nonumber\\
&&\quad  
-\frac{(D-6) (u^2+t^2)}{ t u} \ \xbm2{D}{s}{t}  + {\rm pinchings}\bigg]
\eeqn
Inserting the $\ep$ expansion of $\xbm1{D}{s}{t}$ computed in Ref.~\cite{Bas}
and the $\ep$ expansions of the sub-topologies listed in
Refs.~\cite{Smirnov_Veretin,pentabox,Gonsalves83,Kramer} into
Eq.~(\ref{eq:der_t_xbm1}), and solving it with respect to $\xbm2{D}{s}{t}$,
we obtain , in the physical region $s>0, \ t,u <0$,
\beqn
\label{eq:master2g1g2}
&&\hspace{-0.5cm}\xbm{2}{D}{s}{t} = \Gamma^2(1+\epsilon)
 \Bigg\{
 \frac{G_1(t,u)}{s^3 t} +
 \frac{G_2(t,u)}{s^2 t^2} + \nonumber\\
&&\phantom{\hspace{-0.5cm}\xbm{2}{D}{s}{t} = \Gamma^2(1+\epsilon)
 \Big\{}
 \frac{G_1(u,t)}{s^3 u} +
 \frac{G_2(u,t)}{s^2 u^2}
 \Bigg\}
\, ,\nonumber
\eeqn
where
\begin{eqnarray}
\label{eq:g1tu}
&&\hspace{-0.5cm}
 G_1(t,u) = s^{-2\epsilon} \left\{ \vphantom{\frac{1}{1}} \right.
\frac{6}{{\epsilon}^3}
 + \frac{1}{{\epsilon}^2} \left( 32 - 6\,T - 6\,U \right)
\nonumber \\ && \hspace{0.1cm}
 {}+ \frac{1}{\epsilon}
\left( 1 - 12\,{\pi }^2 - 24\,T + T^2 - 24\,U + 16\,T\,U + U^2 \right)
\nonumber \\ && \hspace{0.1cm}
  -43 - 18\,T + 13\,T^2 + \frac{8}{3}\,T^3
  - 18\,U + 16\,T\,U \nonumber \\ && \hspace{0.1cm}
 {} + 11\,T^2\,U + 13\,U^2 - 20\,T\,U^2 + 
    \frac{8}{3}\,U^3
 + {\pi }^2
     \bigg( 17\,T \nonumber \\ && \hspace{0.1cm}
 {} + 17\,U  -\frac{112}{3} \bigg)
 - 122\,\zeta(3) + 62\,T\,\Li{2}{\mtos}\nonumber \\ && \hspace{0.1cm}
   {}     -  62\,\Li{3}{\mtos}
 + 62\,\snp{1,2}{\mtos}
\nonumber \\ && \hspace{0.1cm}
{}+ i\pi \bigg[
\frac{1}{\epsilon} \left( 16 + 6\,T + 6\,U \right)
  -34 - 9\,{\pi }^2 - 6\,T - 10\,T^2\nonumber \\ && \hspace{0.1cm}
   {}     - 6\,U
    +  14\,T\,U - 10\,U^2
\left. \vphantom{\frac{1}{1}} \bigg] \right\}  ,\nonumber\\
\label{eq:g2tu}
&& \hspace{-0.5cm}
G_2(t,u) = s^{-2\epsilon} \left\{ \vphantom{\frac{1}{1}} \right.
-\frac{2}{{\epsilon}^4}
 + \frac{1}{{\epsilon}^3}
 \left(-8 + \frac{5}{2}\,T + 
     \frac{7}{2}\,U \right) 
\nonumber \\ && \hspace{0.1cm}
{} + \frac{1}{{\epsilon}^2}
 \bigg( -\frac{29}{2} - \frac{5}{12}\,{\pi }^2 + 
     7\,T - T^2 + 20\,U - 4\,T\,U\nonumber \\ && \hspace{0.1cm}
{}  - U^2 \bigg)
 +  \frac{1}{\epsilon}
 \bigg[ -\frac{1}{2}  + 17\,T + 2\,T^2 - 
     \frac{T^3}{3} + \frac{{\pi }^2}{6}
      \big( 14 \nonumber \\ && \hspace{0.1cm}
{} + 5\,T - 29\,U
        \big)  + 13\,U - 28\,T\,U - 4\,U^2
 + 3\,T\,U^2 \nonumber \\ && \hspace{0.1cm}
{} - U^3  + 
     \frac{19}{2}\,\zeta(3) - 2\,T\,\Li{2}{\mtos} + 
     2\,\Li{3}{\mtos}  \nonumber \\ && \hspace{0.1cm}
{}- 2\,\snp{1,2}{\mtos} \bigg]
 + \frac{37}{2} + \frac{37}{40}\,{\pi }^4 + 7\,T - 5\,T^2 \nonumber \\ && \hspace{0.1cm}
{} - \frac{22}{3}\,T^3 + \frac{2}{3}\,T^4 + 5\,U - 20\,T\,U + 
    \frac{8}{3}\,T^3\,U - 2\,U^2
\nonumber \\ && \hspace{0.1cm}
{} + 24\,T\,U^2 - T^2\,U^2 - 
    8\,U^3 - \frac{4}{3}\,T\,U^3 + \frac{4}{3}\,U^4
\nonumber \\ && \hspace{0.1cm}
 {}  +   \frac{{\pi }^2}{6}\left( 79 - 22\,T - 
       5\,T^2 - 200\,U + 76\,T\,U + 
       25\,U^2 \right) \nonumber \\ && \hspace{0.1cm}
{}
  + \left( 68 - 13\,T - 33\,U \right) \,\zeta(3)
 + \Big( 10\,{\pi }^2 - 32\,T + 17\,T^2\nonumber \\ && \hspace{0.1cm}
{} + 12\,T\,U \Big) \,
     \Li{2}{\mtos}
- 36\,\snp{2,2}{\mtos}
\nonumber \\ && \hspace{0.1cm}
{} + \left(28\,T - 6\,U  -32 \right) \,\snp{1,2}{\mtos}
 - 26\,\snp{1,3}{\mtos}\nonumber \\ && \hspace{0.1cm}
{} 
 + \left( 32 - 60\,T - 12\,U \right) \, \Li{3}{\mtos}
 + 86\,\Li{4}{\mtos}
\nonumber \\ && \hspace{0.1cm}
{}+ i\pi \bigg[
\frac{2}{{\epsilon}^3}
 + \frac{1}{{\epsilon}^2}\left( 11 - T + U \right)
 + \frac{1}{\epsilon}
 \bigg( 1 - \frac{31}{6} \,{\pi }^2 - 10\,T\nonumber \\ && \hspace{0.1cm}
{} - 2\,T^2 + 4\,U - 
     2\,T\,U - 2\,U^2 \bigg)
+ 11 + 4\,T - 2\,T^2 \nonumber \\ && \hspace{0.1cm}
{} + \frac{10}{3}\,T^3 + 
    \frac{{\pi }^2}{3} \left( - 65
  + 28\,T - U \right)  + 2\,U - 8\,T\,U \nonumber \\ && \hspace{0.1cm}
{} - 
    8\,U^2
+ 2\,U^3 - 89\,\zeta(3) + 
    \left( 14\,T + 18\,U \right) \,\Li{2}{\mtos} \nonumber \\ && \hspace{0.1cm}
{}  -  32\,\Li{3}{\mtos} + 44\,\snp{1,2}{\mtos}
\vphantom{\frac{1}{1}} \bigg] \bigg\}  ,\nonumber
\end{eqnarray}
$T=\log(-t/s)$, $U=\log(-u/s)$, and where
we used  Nielsen's generalized polylogarithms 
$\mbox{S}_{n,p}$ defined by  
\beqn
&&\hspace{-1cm}\snp{n,p}{x} = \frac{(-1)^{n+p-1}}{(n-1)!\,p!}
\int_0^1 \mbox{d} t \; \frac{ \log^{n-1}(t) \log^p(1-xt) }{t}\nonumber\\
&&\phantom{\hspace{-1cm}\snp{n,p}{x} =}  n,p \ge 1, \quad x \le 1 \nonumber
\eeqn
Expressions for $\xbm{2}{D}{s}{t}$ in the other two kinematic regions, $t>0,
\ s,u<0$ and $u>0, \ s,t<0$ can be easily obtained through the
analytic continuation of the polylogarithms and the logarithms.

It is a strong check of the whole formalism that inserting the analytic
expansion of $\xbm1{D}{s}{t}$ and the derived expression of $\xbm2{D}{s}{t}$
into Eq.~(\ref{eq:der_t_xbm2}), we obtain an equality identically satisfied. 

\vspace{0.5cm}
{\bf Acknowledgements.} \ 
This work has been done in collaboration with C.~Anastasiou, T.~Gehrmann,
E.~Remiddi and J.~B.~Tausk.

\relax
\def\pl#1#2#3{{\it Phys.\ Lett.\ }{\bf #1}\ (#2)\ #3}
\def\zp#1#2#3{{\it Z.\ Phys.\ }{\bf #1}\ (#2)\ #3}
\def\prl#1#2#3{{\it Phys.\ Rev.\ Lett.\ }{\bf #1}\ (#2)\ #3}
\def\rmp#1#2#3{{\it Rev.\ Mod.\ Phys.\ }{\bf#1}\ (#2)\ #3}
\def\prep#1#2#3{{\it Phys.\ Rep.\ }{\bf #1}\ (#2)\ #3}
\def\pr#1#2#3{{\it Phys.\ Rev.\ }{\bf #1}\ (#2)\ #3}
\def\np#1#2#3{{\it Nucl.\ Phys.\ }{\bf #1}\ (#2)\ #3}
\def\sjnp#1#2#3{{\it Sov.\ J.\ Nucl.\ Phys.\ }{\bf #1}\ (#2)\ #3}
\def\app#1#2#3{{\it Acta Phys.\ Polon.\ }{\bf #1}\ (#2)\ #3}
\def\jmp#1#2#3{{\it J.\ Math.\ Phys.\ }{\bf #1}\ (#2)\ #3}
\def\jp#1#2#3{{\it J.\ Phys.\ }{\bf #1}\ (#2)\ #3}
\def\nc#1#2#3{{\it Nuovo Cim.\ }{\bf #1}\ (#2)\ #3}
\def\lnc#1#2#3{{\it Lett.\ Nuovo Cim.\ }{\bf #1}\ (#2)\ #3}
\def\ptp#1#2#3{{\it Progr. Theor. Phys.\ }{\bf #1}\ (#2)\ #3}
\def\tmf#1#2#3{{\it Teor.\ Mat.\ Fiz.\ }{\bf #1}\ (#2)\ #3}
\def\tmp#1#2#3{{\it Theor.\ Math.\ Phys.\ }{\bf #1}\ (#2)\ #3}
\def\jhep#1#2#3{{\it J.\ High\ Energy\ Phys.\ }{\bf #1}\ (#2)\ #3}
\def\epj#1#2#3{{\it Eur.\ Phys. J.\ }{\bf #1}\ (#2)\ #3}
\relax


\end{document}